\documentclass[printer]{aa}
\usepackage{natbib}
\usepackage{url}
\usepackage{amsmath}
\usepackage{graphicx}
\usepackage{aas_macros}

\begin{document}

  \title{3DEX: a code for fast spherical Fourier-Bessel \\ decomposition of  3D surveys}
\author{B. Leistedt \inst{1,4} \thanks{boris.leistedt@gmail.com} \and A. Rassat \inst{2,4}\thanks{anais.rassat@epfl.ch} \and A. R\'efr\'egier \inst{3} \and J.-L. Starck \inst{4}}
\institute{ $^1$ Department of Physics and Astronomy, University College London, London WC1E 6BT, United Kingdom.  \\
$^2$ Laboratoire d'Astrophysique, Ecole Polytechnique F\'ed\'erale de Lausanne (EPFL), Observatoire de Sauverny, CH-1290, Versoix, Switzerland.\\ 
 $^3$ Institute for Astronomy, ETH Z\"urich, Wolfgang-Pauli-Strasse 27, CH-8093 Z\"urich, Swtitzerland.\\
 $^4$ Laboratoire AIM, UMR CEA-CNRS-Paris 7, Irfu, SAp/SEDI, Service d'Astrophysique, CEA Saclay, F-91191 GIF-SUR-YVETTE CEDEX, France.}


\abstract
{High-precision cosmology requires the analysis of large-scale surveys in 3D spherical coordinates, i.e. spherical Fourier-Bessel decomposition.  Current methods are insufficient for future data-sets from wide-field cosmology surveys.}
{The aim of this paper is to present a public code for fast spherical Fourier-Bessel decomposition that can be applied to cosmological data or 3D data in spherical coordinates in other scientific fields. }
{We present an equivalent formulation of the spherical Fourier-Bessel decomposition that separates radial and tangential calculations.  We propose the use of the existing pixelisation scheme HEALPix for a rapid calculation of the tangential modes. }
{3DEX (3D EXpansions) is a public code for fast spherical Fourier-Bessel decomposition of 3D all-sky surveys that takes advantage of HEALPix for the calculation of tangential modes. We perform tests on very large simulations and we compare the precision and computation time of our method with an optimised implementation of the spherical Fourier-Bessel original formulation. For surveys with millions of galaxies, computation time is reduced by a factor 4-12 depending on the desired scales and accuracy. The formulation is also suitable for pre-calculations and external storage of the spherical harmonics, which allows for additional speed improvements. The 3DEX code can accommodate data with masked regions of missing data. 3DEX can also be used in other disciplines, where 3D data are to be analysed in spherical coordinates. The code and documentation can be downloaded at \url{http://ixkael.com/blog/3dex}.}
{}

\maketitle

\section{Introduction}\label{sec:introduction}

In the last few decades, cosmology has become a data-driven field, where high-precision measurements of the cosmic microwave background \citep[CMB, e.g.,][]{wmap7}, weak lensing \citep[e.g.,][]{Schrabback:2010} and galaxy surveys \citep[e.g.,][]{Percival:2007} have permitted the establishment of a standard cosmological model in which the Universe is composed of 4\% baryons, 22\% dark matter and 74\% dark energy.   Some major questions remain, the nature of dark matter and dark energy in particular is still not understood. Similarly, the initial conditions of the Universe are yet to be established and alternative models of gravity are still to be tested in comparison with Einstein's general relativity. 

New surveys are underway with these science objectives, e.g. Planck for the CMB \citep{Planck}, DES \citep[Dark Energy Survey,][]{DES:2005}, BOSS \citep[Baryon Oscillation Spectroscopic Survey,][]{Schlegel:2007}, LSST \citep[Large Synoptic Survey Telescope,][]{LSST} and Euclid \citep{2011arXiv1110.3193L, Euclidsb} for weak lensing and the study of large-scale structure with galaxy surveys.  In order to be beneficial, cosmological studies of these surveys need to use high-precision statistical methods, such as a full 3D analysis on the sky where all-sky 3D surveys are available.

Several tools have been developed to analyse data on the sphere, which is required for a 2D spherical harmonic CMB analysis \citep{igloo,igloo2,healpix:2002,gorski:2004by,glesp}. Weak lensing and galaxy survey data can also be analysed tomographically (i.e. in 2D slices), but unlike for the CMB, a full 3D spherical Fourier-Bessel analysis can also be sought \citep{Fisher:1995,Heavens:1995,heavens3d,weak3d,2011arXiv1112.3100R}. Previous 3D data analyses were on relatively small data sets \citep{Fisher:1995,Heavens:1995,Erdogdu:2006dv,Erdogdu:2005wi}, but future surveys like Euclid and LSST will provide surveys with billions of galaxies, making previous methods for calculating the 3D spectra unfeasibly time-consuming.

In Section \ref{sub:FB}, we present the theory behind the 3D Fourier-Bessel decomposition for infinite and finite continuous fields as well as the usual method for a discrete survey (e.g. galaxy survey). In section \ref{sub:3equiv}, we present two additional equivalent formulations of the spherical Fourier-Bessel decomposition, one of which is central to the 3DEX code.  In Section \ref{sec:Methods Comparison}, we compare the accuracy and calculation time for the usual method used for calculating Fourier-Bessel coefficients and methods with the 3DEX code presented in this paper. In Section \ref{sec:3DEX}, we describe the 3DEX library and give examples of how to use it. In Section \ref{sec:Conclusion} we present our conclusions. We also include an appendix, where we discuss the subtleties of the Fourier-Bessel normalisation.

\section{Theory}\label{sec:theory}
\label{sec:Mathematical Approach}

\subsection{The spherical Fourier-Bessel decomposition}\label{sub:FB}

In observational cosmology, spherical coordinates (where the observer is at the origin) are a natural choice for the analysis of cosmological fields. In this system of coordinates, eigenfunctions of the Laplacian operator are products of spherical Bessel functions and spherical harmonics, i.e. functions $j_\ell (kr)Y_{\ell m}$ with eigenvalues $-k^2$. For an homogeneous three-dimensional  field $f(\textbf{r}) = f(r,\theta,\phi)$ in a flat geometry, the spherical Fourier-Bessel decomposition \citep{Fisher:1995,heavens3d,weak3d} is
\begin{eqnarray}
	\hspace*{-0mm} f(r,\theta,\phi) =  \sqrt{\frac{2}{\pi}}\int dk \sum_{\ell m} f_{\ell m}(k) k j_\ell (kr) Y_{\ell m}(\theta,\phi) ,\label{thraw1}
\end{eqnarray}
with the inverse relation
\begin{eqnarray}
	f_{\ell m}(k) = \sqrt{\frac{2}{\pi}} \int d^3\textbf{r}\ f(r,\theta,\phi)k j_\ell (kr)Y^{*}_{\ell m}(\theta,\phi) .\label{thraw2} 
\end{eqnarray}
{Note that this decomposition uses the same notation as \cite{2011arXiv1112.3100R} and \cite{weak3d}, which is slightly different from the one used in \cite*{2011arXiv1112.0561L}.}
The coefficients may be used to calculate the 3D power spectrum $C(\ell, k)$, defined by
\begin{equation} \left< f_{\ell m}(k)f^*_{\ell' m'}(k')\right>=C(l,k) \delta_D(k-k')\delta_{\ell \ell'}\delta_{mm'},\end{equation} 
a na\"\i ve estimator of which is
\begin{eqnarray}
	C_\ell (k) = \frac{1}{2l+1} \sum_m |f_{\ell m}(k)|^2.
\end{eqnarray}
This can be seen as an extension of the usual 2D power spectrum $\left< f_{\ell m}f^*_{\ell' m'}\right>=C_l \delta_{\ell \ell'}\delta_{mm'}$. The latter arises from the spherical harmonic transform of a 2D field given on the sphere $f(\theta,\phi) = \sum_{\ell m}f_{\ell m}Y_{\ell m}(\theta,\phi)$.

In practice, surveys will only cover a finite amount of volume, limiting the analysis to a sphere of radius $R$. These boundary conditions lead to a discrete spectrum $\{ k_{\ell n} \}$, which is detailed in the appendices. In this paper, we assumed as a boundary condition that $f$ vanishes at $r=R$. The spherical Fourier-Bessel decomposition becomes \citep{Erdogdu:2006dv,Fisher:1995}
\begin{eqnarray}
	\hspace*{-0mm}f(r,\theta,\phi) = \sum_{\ell mn} \kappa_{\ell n} f_{\ell m}(k_{\ell n}) k_{\ell n} j_\ell (k_{\ell n}r)Y_{\ell m}(\theta,\phi), \label{fbreconstr}
\end{eqnarray}
which is exact if the ranges of $\ell$,$m$ and $n$ are infinite. The Fourier-Bessel coefficients are denoted by $f_{\ell mn} = f_{\ell m}(k_{\ell n})$, and $\kappa_{\ell n}$ is the normalisation constant (see appendices for more details).

In various applications, though, the continuous field $f$ cannot  be directly observed. This is notably the case in cosmology where galaxy surveys give indirect information about the underlying matter density field through their spacial positions. Note that these tracers are subject to various distortions and non-linearities, but these are not the purpose of this work. In this work we only consider linear or quasi-linear scales ($\ell< 50$,  $~k<0.2 {\rm hMpc}^{-1}$).

If the only information about the field $f$ is a list of coordinates $\textbf{r}_p = (r_p,\theta_p,\phi_p)$ with $p=1,\dots,N$ (where $N$ is the number of galaxies in the latter example), the survey may be considered as a superposition of 3D Dirac deltas and each coefficient $f_{\ell mn}$ can simply be estimated with a sum \citep{Heavens:1995,Fisher:1995,Erdogdu:2006dv,cmbbox} 
\begin{eqnarray}
	\tilde{f}(\textbf{r}) & = & \sum^N_{p=1} \delta^{(3)}(\textbf{r}-\textbf{r}_p), \\
	\tilde{f}_{\ell mn} & = & \sum^N_{p=1} k_{\ell n}j_\ell (k_{\ell n}r_p)Y^{*}_{\ell m}(\theta_p,\phi_p). \label{rawestimate}
\end{eqnarray}

This formulation has been used for the analysis of shallow galaxy surveys such as the IRAS 1.2mJ survey \citep[$\sim6k$ galaxies,][]{IRAS,Fisher:1995,Heavens:1995}, and the 2MRS survey \citep[2MASS Redshift Survey, $\sim 45$k galaxies,][]{2MRS,Erdogdu:2006dv,Erdogdu:2005wi}. Since the time to calculate equation \ref{rawestimate} is proportional to $N n_{\rm max}(\ell_{\rm max}+1)^2/2$, Equation \ref{rawestimate} will become highly time-consuming when applied to larger surveys or when precise decomposition is required (large $n_{max}$ and $\ell_{max}$).

\subsection{Three equivalent formulations}\label{sub:3equiv}

In spherical coordinates, since 3D space can be viewed as an infinite series of closed shells $\Omega(r)$, the spherical Fourier-Bessel decomposition may also arise from repeated 2D spherical harmonic transforms to which spherical Bessel transforms are applied \citep{cmbbox}. Formally, the field $f$ given on each shell $\Omega(r)$ is first expanded into spherical harmonics
\begin{eqnarray}
	f(r,\theta,\phi) = \sum_{\ell m} f_{\ell m}(r) Y_{\ell m}(\theta,\phi) ,\label{thdirect1}
\end{eqnarray}
for which the inversion formula gives harmonics coefficients $f_{\ell m}(r)$ depending on the radius $r$
\begin{eqnarray}
	f_{\ell m}(r) = \int_{\Omega(r)} d\Omega \ f(r,\theta,\phi) Y^{*}_{\ell m}(\theta,\phi). \label{thdirect2}
\end{eqnarray}
It is then possible to perform a spherical Bessel transform
\begin{eqnarray}
	f_{\ell m}(r) = \sqrt{\frac{2}{\pi}} \int dk \  f_{\ell m}(k) kj_\ell (kr), \label{thdirect3}
\end{eqnarray}
leading to the final Fourier-Bessel coefficients $f_{\ell m}(k)$
\begin{eqnarray}
	f_{\ell m}(k) = \sqrt{\frac{2}{\pi}} \int dr \ r^2 f_{\ell m}(r) kj_\ell (kr) .\label{thdirect4} 
\end{eqnarray}
This formulation hence extends the notion of 2D spherical harmonics to three-dimensional fields. 

It is also possible to conceive the reverse approach, i.e. to perform the spherical Bessel transform first and subsequently expand the resulting coefficients into spherical harmonics. Formally, the $\ell$-th order spherical Bessel transform of $f$ (similar to its Hankel transform) is
\begin{eqnarray}
	f(r,\theta,\phi) =  \sqrt{\frac{2}{\pi}} \int dk \  f_{\ell }(k,\theta,\phi) kj_\ell (kr) ,\label{threverse1}
\end{eqnarray}
for which the inversion formula gives
\begin{eqnarray}
	f_\ell (k,\theta,\phi) =  \sqrt{\frac{2}{\pi}}  \int dr \ r^2 f(r,\theta,\phi) kj_\ell (kr) .\label{threverse2}
\end{eqnarray}
The result is then expanded into spherical harmonics but with an unusual formulation since $f_\ell (k,\theta,\phi)$ and $Y_{\ell m}(\theta,\phi)$ (as well as the basis functions $j_\ell (kr)$ and $Y_{\ell m}(\theta,\phi)$) have now the $\ell$ parameter in common:
\begin{eqnarray}
	f_\ell (k,\theta,\phi) =  \sum_{m} f_{\ell m}(k) Y_{\ell m}(\theta,\phi) .\label{threverse3}
\end{eqnarray}
Again, using the inversion formula, we obtain the Fourier-Bessel coefficients
\begin{eqnarray}
	f_{\ell m}(k) =  \int_{\Omega} d\Omega \ f_\ell (k,\theta,\phi) Y^{*}_{\ell m}(\theta,\phi). \label{threverse4}
\end{eqnarray}
Due to the closed domains of shells $\Omega(r)$ and thus the relative independence of angular and radial dimensions, the \emph{raw} (equations \ref{thraw1} and \ref{thraw2}), the \emph{forward} (denoted by SHB for \emph{spherical-Harmonic-Bessel}, equations \ref{thdirect1} to \ref{thdirect4}) and the \emph{reverse} (denoted by SBH for \emph{spherical-Bessel-Harmonic}, equations \ref{threverse1} to \ref{threverse4}) methods are equivalent formulations of the spherical Fourier-Bessel decomposition of any three-dimensional field $f(r,\theta,\phi)$.  This is summarised in the following schematic description of each method:  
\begin{eqnarray}
\textnormal{RAW : } \ & & f(\textbf{r}) \  \xrightarrow[]{\textrm{three-dimensional integration}} \ f_{\ell m}(k)
\nonumber \\
\textnormal{SHB : } \ & & f(\textbf{r}) \ \xrightarrow[]{\textrm{SHT}}  \ \ \ \ f_{\ell m}(r) \ \ \ \ \xrightarrow[]{\textrm{SBT}} \ \ f_{\ell m}(k)
\\
\textnormal{SBH : } \ & & f(\textbf{r}) \ \xrightarrow[]{\textrm{SBT}} \ \ f_{\ell }(k,\theta,\phi) \ \ \xrightarrow[]{\textrm{SHT}} \ \ f_{\ell m}(k) \nonumber.
\end{eqnarray}

Note that this section is related to the ideal case $R = \infty$, but all equations can be straightforwardly rewritten for a finite $R$ by replacing $k$ by $k_{ln}$, bounding each integral and adapting normalisation. The formulas arising from this adaptation are used in the next sections.

\subsection{Estimating Fourier-Bessel coefficients from a real survey}\label{sub:estfbd}

Although the three approaches described in \ref{sub:3equiv} are theoretically equivalent, their estimates and numerical implementations take different forms.

\subsubsection{\emph{Forward} approach (SHB)}
Estimating the $f_{\ell mn}$ coefficients using the \emph{forward} method naturally requires the radial dimension to be discretised. Indeed, the first step is to compute the spherical harmonic transform on a set of shells located at radial values $r_1, \dots, r_{N_{layers}}$. In each layer, the coefficients $f_{\ell m}(r_i)$ are estimated. Although it is possible to perform a raw estimate for the later harmonics transform, it is often advisable to use a robust 2D discretisation scheme (of $N_{pix}(i)$ pixels for the $i$-th shell) and to take advantage of the related high-performance algorithms. Angular space is hence discretised into nodes $(r_i,\theta_p,\phi_p)=(r_i,{\boldsymbol{\gamma}}_q)$ and the field is approximated on each node, giving $\tilde{f}(r_i,{\boldsymbol{\gamma}}_p)$. The spherical harmonic decomposition in the $i$-th shell becomes
\begin{eqnarray}
	\tilde{f}_{\ell m}(r_i) = \sum^{Npix(i)}_{p=1} \tilde{f}(r_i,{\boldsymbol{\gamma}}_p)Y^{*}_{\ell m}({\boldsymbol{\gamma}}_p),
\end{eqnarray}
and the final coefficients are obtained by performing the following spherical Bessel decomposition:
\begin{eqnarray}
 \tilde{f}_{\ell mn} =  \sum_{i=1}^{N_{layers}}\tilde{f}_{\ell m}(r_i) k_{\ell n}j_\ell (k_{\ell n}r_i).
\end{eqnarray}
With this method, radial and angular spaces are discretised and both transforms are approximated.

\subsubsection{\emph{Reverse} approach (SBH)}
For the \emph{reverse} approach, a 2D scheme on the sphere was required as well. As previously, this scheme defines a set of $N_{pix}$ zones (pixels) related to angular nodes $\boldsymbol{\gamma}_q$. If $G_q$ denotes the points of the survey located in the solid angle corresponding to the $q$-th zone of the scheme, we perform the spherical Bessel Transform (raw estimate) in each zone
\begin{eqnarray}
 	\tilde{f}_{\ell n}({\boldsymbol{\gamma}}_q) = \tilde{f}_\ell (k_{\ell n},{\boldsymbol{\gamma}}_q) =  \sum_{p \in G_q} k_{\ell n}j_\ell (k_{\ell n}r_p), \label{revdiscr1}
\end{eqnarray}
and each of these intermediate maps is decomposed into spherical harmonic (spherical Harmonics Transform) which gives the Fourier-Bessel coefficients
\begin{eqnarray}
	\tilde{f}_{\ell mn} = \sum^{Npix}_{q=1} \tilde{f}_{\ell n}({\boldsymbol{\gamma}}_q)Y^{*}_{\ell m}({\boldsymbol{\gamma}}_q) \label{revdiscr2}
.\end{eqnarray}

With the reverse method, one can avoid to discretise radial space. Moreover, this one-shell pixelisation of the sky (thus based on physical solid angles) allows for a natural treatment of radial distortions (redshift, relativistic) and masking effects. Using multiple resolutions at different radial values, as would be possible with the forward method, is much more questionable. {The SHB method also proves to be a powerful tool for weighting the data prior to estimating the power spectrum. For instance, in \cite{1999MNRAS.305..527T} used a fiducial power spectrum to derive an optimal weighting operation. This operation is quite complex when using the raw Fourier-Bessel approach, whereas the SHB formulation naturally handles the dependence on $k$ of the weighting function.}

The three methods to estimate the spherical Fourier-Bessel decomposition can therefore also be expressed for a discrete 3D survey, summarised schematically below:

\begin{eqnarray}
\textnormal{RAW : } &\{\textbf{r}_p\}  \xrightarrow[]{\textrm{Raw sum, best estimate of FB coefficients}}  \ \tilde{f}_{\ell mn}
\nonumber \\
\textnormal{SHB : } & \{\textbf{r}_p\} \xrightarrow[]{\textrm{Approx SHT}}  \ \tilde{f}_{\ell m}(r_i) \ \ \xrightarrow[]{\textrm{Approx SBT}} \ \ \tilde{f}_{\ell mn}\nonumber \\
\textnormal{SBH : } & \{\textbf{r}_p\} \xrightarrow[]{\textrm{Exact SBT}} \ \ \ \tilde{f}_{\ell n}(\boldsymbol{\gamma}_p) \ \ \xrightarrow[]{\textrm{Approx SHT}} \ \tilde{f}_{\ell mn} \nonumber.
\end{eqnarray}
 
Note that in practice, the range of $(l,m,n)$ is finite, which introduces an additional approximation. Here, $\ell$ and $n$ are restricted to $[0,\ell _{max}]$ and $[1,n_{max}]$ respectively. Given $\ell$, $m$ goes from $-\ell$ to $\ell$.

\section{Method comparison}
\label{sec:Methods Comparison}

\subsection{Complexity, accuracy and discretisation grids}

For a survey that probes a field by $N$ discrete points, the raw method is the natural estimate of the Fourier-Bessel coefficients. However, since each point contributes to the calculation of every coefficient $\tilde{f}_{\ell mn}$ ($\forall \ l,m,n$), computation time is proportional to $N \cdot n_{max} ( \ell _{max} + 1)^2/2$, which can be highly problematic for large surveys. 

In the forward method, the repeated spherical harmonic transforms take advantage of tesselation schemes and high-performance algorithms such as those provided by HEALPix [\cite{gorski:2004by}], IGLOO [\cite{igloo}] or GLESP [\cite{glesp}]. Roughly speaking, the number of nodes to be considered is reduced from $N$ to $N_{pix}$, and the use of fast spherical harmonic transforms on these schemes significantly decreases computation time. 

However, this approach requires the three-dimensional space to be divided into shells $\Omega(r_i)$. Both radial and angular dimensions are discretised, and the survey is approximated on an actual 3D grid. In practice, this approximation deteriorates the accuracy of the estimated Fourier-Bessel coefficients. Furthermore, designing a meaningful radial discretisation is a difficult task. For equal-area pixelisations, the area of each pixel on the $i$-th shell is $4\pi r_i^2/N_{pix}(i)$. With HEALPix, the $n_{side}$ angular parameter may only be increased by a factor 2, which changes the number of pixels by a factor of 4 (since $N_{pix}(i)=12 n_{side}(i)^2$).  This means that pixel areas cannot be stabilised for subsequent shells as $r$ increases. Consequently, it is not possible to adapt a resolution to obtain a 3D scheme with equal-volume voxels. Extending 2D schemes to 3D is difficult and may even require a novel formalism for an equal-voxel 3D grid.

In the reverse approach, though, the use of angular 2D schemes is possible, but radial space does not need to be discretised. The points of the survey are grouped according to angular zones instead of being approximated on a 3D grid. An estimate of the spherical Bessel transform is computed in every solid angle, and the result is then expanded in spherical harmonics on the 2D spherical grid. In the final account, this method naturally leads to more accurate coefficients than the forward method and also takes advantage of high-performance 2D schemes. For these reasons, we focus on the reverse approach and its implementation, using HEALPix for the angular transform.

Finally, for both forward and reverse methods the spherical harmonics discretised basis (coefficients $Y_{\ell m}(\gamma_p)$) may be fully pre-computed and stored in external files. This is a particularly useful feature (incompatible with the original formulation of spherical Fourier-Bessel), which significantly eases and speeds up the use of these methods.

\subsection{Speed and accuracy of the Reverse Method}
To test the accuracy and speed of the reverse method compared to the raw method, we considered the high-resolution full-sky Horizon simulation \citep{Horizon}.  Horizon is a N-body simulation covering a 2$h^{-1}{\rm Gpc}$ periodic box using 70 billion dark matter particles using a WMAP3 cosmology \citep{Spergel:2006hy}. A derived halo catalogue is available, which we used to calculate $f_{\ell mn}$ and $C_\ell(k_{\ell n})$ values using both methods (raw and reverse). Since we are interested only in comparing the speed of each method, we simply considered each halo to have equal weight.

We performed the raw and the reverse estimates on three `surveys' of $N=4.2\times 10^5, 3.1 \times10^6$ and $1\times 10^7$ halos, which correspond to three different depths ($z_{max}=0.1, 0.2$ and $0.3$ respectively) in the Horizon simulation. The HEALPix angular parameter is given by $n_{side}$.

The results of the accuracy and speed tests are given in Table 1. The third (fourth) column gives the percentage f coefficients for which the relative accuracy $\epsilon(f_{\ell m n})$ ($\epsilon(C_{\ell n})$) is lower than 0.3\% for given values of $n_{side}$ and $N$.   We considered the intervals $(l,n) \in ([0,20],[1,20])$ and $(l,n) \in ([21,50],[21,50])$ separately, since the estimation of higher coefficients depends more on the value of $n_{side}$.  We also compared computation times of the two methods by observing the ratio $T = t_{reverse}/t_{raw}$. Given a survey and a method, computation time denotes the CPU time required to compute the $k_{ln}$'s (from the Bessel functions) and the final coefficients $f_{\ell mn}$ without using pre-computed quantities. Note that we performed this analysis by distributing the calculations on five machines and simply adding the individual contributions to computation time since our method is linear with survey size. With the reverse method, though, the roots of the Bessel functions as well as the spherical harmonics may be pre-computed and stored in external files, which decreases computation time and complexity when working with 3DEX. 

\begin{table}[htbp]
	\begin{center}
    	\begin{tabular}{ cccccc }
\hline
 $N$	&$n_{side}$ 	&	 $\epsilon_r(f_{\ell mn}) < 0.3\%$ 	&	 $\epsilon_r(C_{\ell n}) < 0.3\%$   & T	\\
 	&		&	 [0,20] / [21,50]	& [0,20] / [21,50] \\
\hline
4.2e5	&	512	&	87\% / 42\%	&	99\% / 96\%	&	8	\\
	&	1024	&	95\% / 65\%	&	99\% / 98\%	&	4	\\
	&	2048	&	99\% / 84\%	&	99\% / 99\%	&	2	\\
3.1e6	&	512	&	92\% / 50\%	&	99\% / 95\%	&	10	\\
	&	1024	&	98\% / 74\%	&	100\% / 100\%	&	5	\\
	&	2048	&	99\% / 90\%	&	100\% / 100\%	&	2	\\
9.7e6	&	512	&	92\% / 50\%	&	100\% / 97\%	&	12	\\
	&	1024	&	97\% / 74\%	&	100\% / 100\%	&	6	\\
	&	2048	&	99\% / 90\%	&	100\% / 100\%	&	3	\\
\hline \\
	\end{tabular}
	\caption{Estimation of Fourier-Bessel coefficients: comparison of the new method, the \textit{reverse} formulation (equations \ref{revdiscr1} and \ref{revdiscr2} using HEALPix discretisation) with the original, \textit{raw} formulation (equation \ref{rawestimate}). The third (fourth) column gives the percentage f coefficients for which the relative accuracy $\epsilon(f_{\ell m n})$ ($\epsilon(C_{\ell n})$) is lower than 0.3\% for given values of $n_{side}$ and $N$. T is the ratio of elapsed times of the two methods.}
 	\end{center}
\end{table}

The reverse method is about an order of magnitude faster than the raw method, but this depends on the choice of $n_{side}$. For $n_{side}=1024$ almost all $f_{\ell mn}$ coefficients in the range $[0,20]$ (for $\ell$ and $n$) have relative error below $1\%$, and $90\%$ have it below $0.3\%$, whereas over 99\% of $C(\ell,k_n)$ coefficients are accurate to $<0.3\%$.  In the range $[20,50]$, the accuracy is somewhat degraded due to the extension of the HEALPix formalism to 3D surveys. Indeed, for data distributed on the sphere, 3D space is very sparse even for large surveys. Increasing $n_{side}$ to 1024 or 2048 strongly improves the accuracy for higher orders $\ell$. Note that comparisons for $\ell >50$ are limited by the amount of time the raw method takes.

Figures 1 and 2 show the time taken for calculations as a function of $\ell_{max}$ and $n_{max}$ (Figure 1), and as a function of number of halos (Figure 2).  The boxes correspond to the raw method, the circles and diamonds to the reverse method with $n_{side}=512, 1024$ respectively. The dashed line corresponds to the general rule that the raw method scales as $N n_{max} \left(\ell_{max}+1\right)^2/2$, whereas the points are all estimated from calculations. With $\ell _{max}=n_{max}=100$ and $N=9.7e6$, the raw decomposition took a few days of calculations, whereas the reverse method only took 12 hours. {In our formalism, $k_{max}$ is determined by the choice of $R$ for the boundary condition and by the band-limit $n_{max}$ for spherical Bessel coefficients. For each multipole $\ell$ we have $k_{max}=k_{\ell n_{max}} = q_{\ell n_{max}}/R$ where $q_{\ell n_{max}}$ is the $n_{max}$-the root of the $\ell$-th spherical Bessel function. Because $R$ is usually imposed by the problem or the data, one must increase $n_{max}$ to probe smaller radial scales. In fact, a reasonable approximation (or even a simple plot) shows that $q_{\ell n} \approx (\ell + 3n)$. This observation enabled to predict which radial scales are probed and how computation time scales with $k_{max}$, given that we provide the complexity for $\ell_{max}$ and $n_{max}$.}

\begin{figure}
	\centering
	\hspace*{-3mm}
	\includegraphics[width=9.2cm]{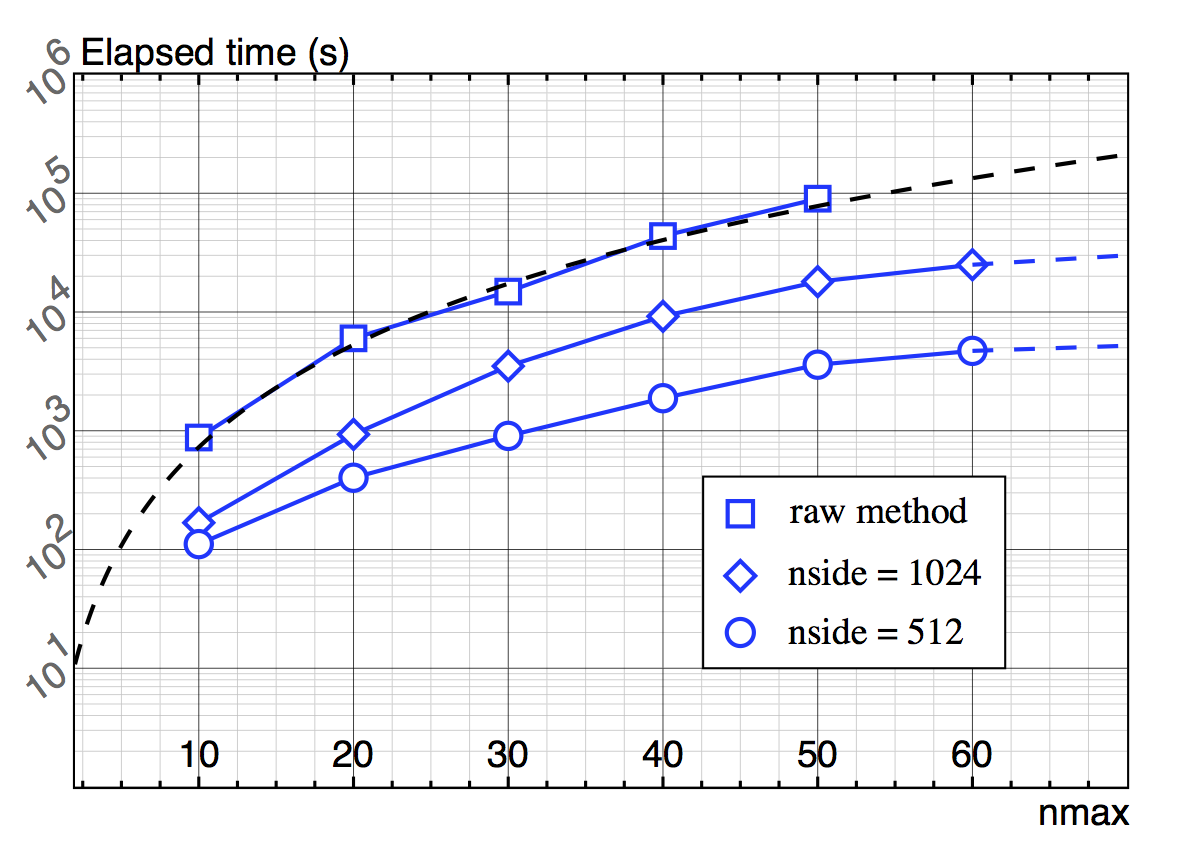}\label{fig:increasingL}
	\begin{minipage}{8.2cm}\caption{Speed results (raw and reverse methods) for increasing summation limits $\ell _{max}=n_{max}$, for a survey of $N=9.7 \times 10^6$ halos. Dashed lines are the fitted complexity curves. The reverse formulation is suitable for pre-calculations and external storage of the spherical harmonics, which was not performed here but enables for additional speed improvements.}\end{minipage}
	\label{fig1}
\end{figure}
\begin{figure}
	\centering
	\hspace*{-3mm}
	\includegraphics[width=9.2cm]{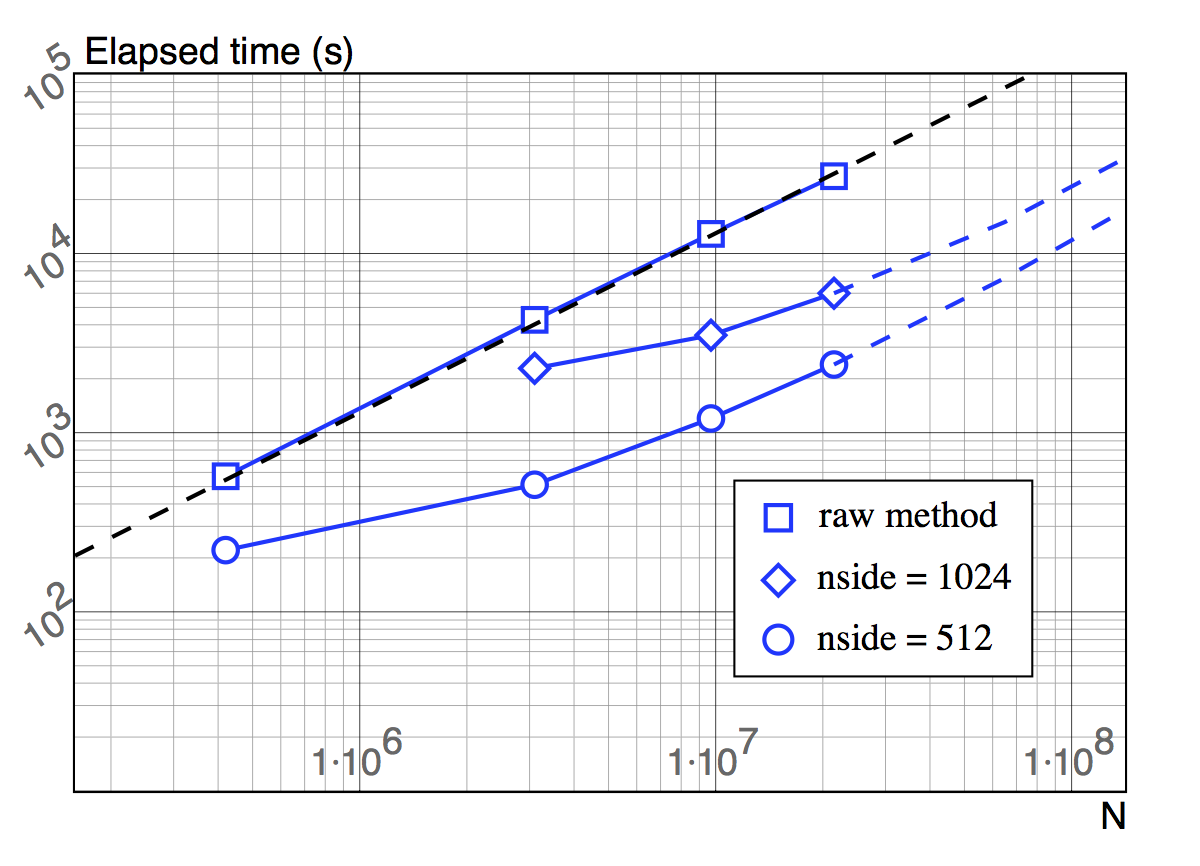}\label{fig:increasingN}
	\begin{minipage}{8.2cm}\caption{Speed results (raw and  reverse methods) for increasing survey size, for $\ell _{max}=n_{max}=30$. Dashed lines are the fitted complexity curves.}\end{minipage}
	\label{fig2}
\end{figure}

One of the main advantages of the reverse method is that it is naturally suited to parallel computing because it uses HEALPix fast spherical Harmonics Transform routines. All previous tests were performed on a recent computer (i7 processor, 8Go RAM) and take advantage of OpenMP (with four threads). More advanced computing means (larger RAM and more processors) significantly decrease calculation time. For example, $\ell _{max}=n_{max}=128$ with $n_{side}=2048$ took about an hour with 128 cores and 512Go RAM, whereas computation time for the raw method was estimated to several days on the same machine. Note that the raw method is also suited to parallelisation: galaxies may be treated separately by different threads. In all experiments, we took advantage of this property and performed both \textit{raw} and \textit{reverse} decompositions with four threads to perform relevant comparisons between the two.

\begin{figure*}
	\centering
	\hspace*{-9mm}\resizebox{\hsize}{!}{ 
		\includegraphics[width=8.5cm]{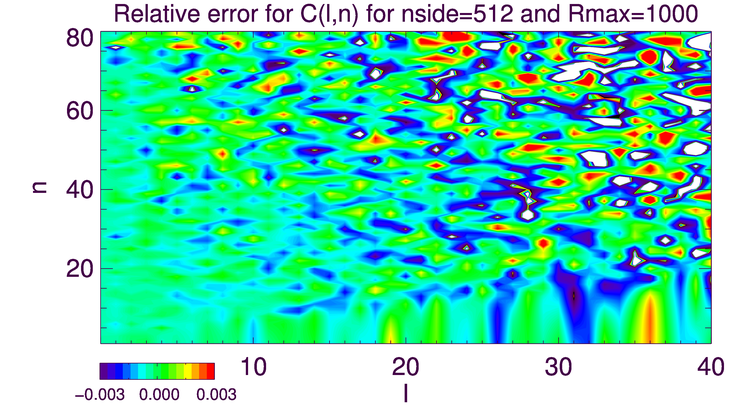}
		\includegraphics[width=8.5cm]{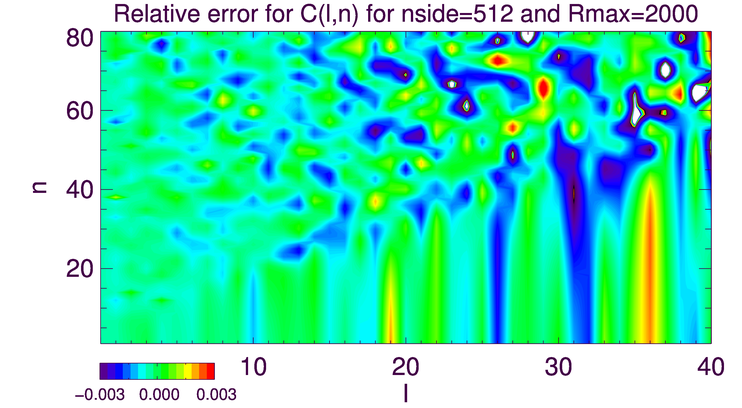}\quad}\\
	\hspace*{-9mm}\resizebox{\hsize}{!}{\includegraphics[width=8.5cm]{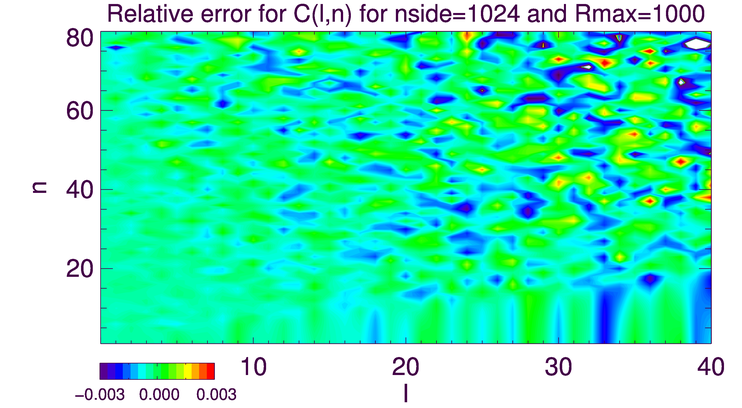} 
		\includegraphics[width=8.5cm]{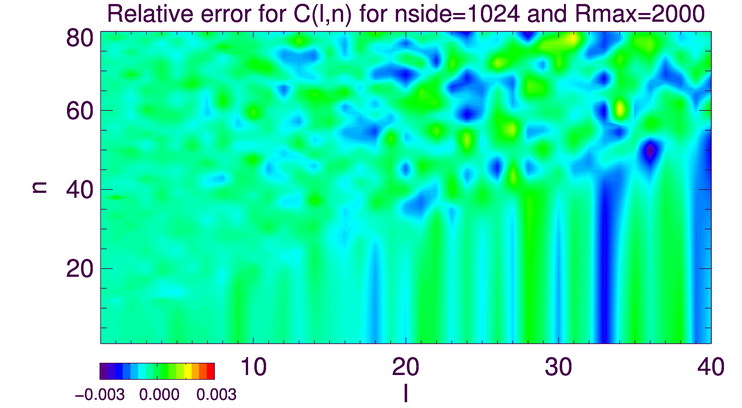}\quad}\\
	\parbox[b]{15cm}{
		\caption{Relative error on the power spectrum in mode-mode space $C(l,n)$. We compare the original formulation of spherical Fourier-Bessel decomposition with the reverse formulation, testing $n_{side}=512, 1024$ (rows) and $R=1000, 2000$ (columns). Only a few zones (white spots) are outside the range $[-0.3\%,+0.3\%]$.}
		\label{fig:nspace}
		}
\end{figure*}

In terms of the power spectrum, figures 3 and 4 show the relative error between the raw and the reverse methods both in mode-mode space ($\ell - n$) and in mode-scale space ($\ell - k_{\ell n}$). For this comparison we decomposed a survey of $N= 4.2\times 10^5$ halos with $z_{max}=0.1$. Figures in mode-mode space naturally differ according to the choice of the boundary $R$ because the latter determines the discrete radial scale spectrum $\{ k_{\ell n} \}$, and hence mode $n$ computed with two different $R$'s corresponds to different $k$-scales. When comparing the results from $R=1000$ and $R=2000$ in mode-scale space, we observe that the boundary condition fixes the explored scales. The left column is thus complementary to the right column to explore higher values of $k$. Although figure \ref{fig:nspace} gives information about the final coefficients, figure 4 is hence more appropriate to see which scales are probed and with what accuracy.

In view of the $\ell - k_{\ell n}$ space, we see that no fluctuations are observed along the $k$ axis up to $k=0.03 {\rm hMpc^{-1}}$. In this range, fluctuations occur in $\ell$ space, which are accurately probed with $n_{side}=512$ until $\ell=25$ but naturally require a more precise scheme for $\ell >25$, $k>0.03 {\rm hMpc^{-1}}$ (smaller scales in physical space). In conclusion, parameter $n_{side}$ (as well as $R$) must be chosen depending on the scales one wishes to probe. Figures 3 and 4 provide accuracy results that are complementary to Table 1.

\begin{figure*}
	\centering
	\hspace*{-9mm}\resizebox{\hsize}{!}{
		\includegraphics[width=8.5cm]{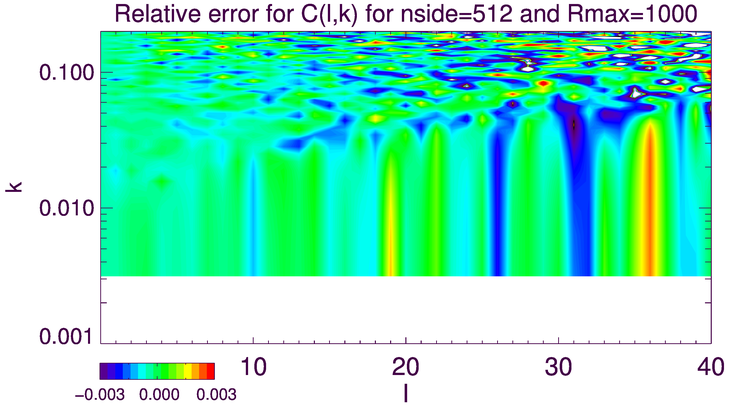}
		\includegraphics[width=8.5cm]{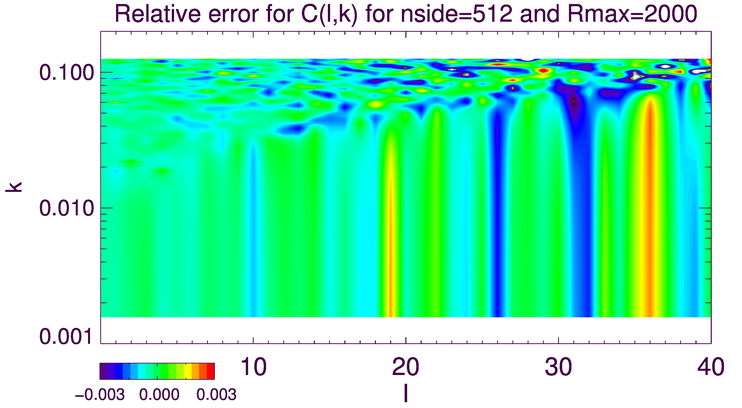}\quad}\\
	\hspace*{-9mm}\resizebox{\hsize}{!}{
		\includegraphics[width=8.5cm]{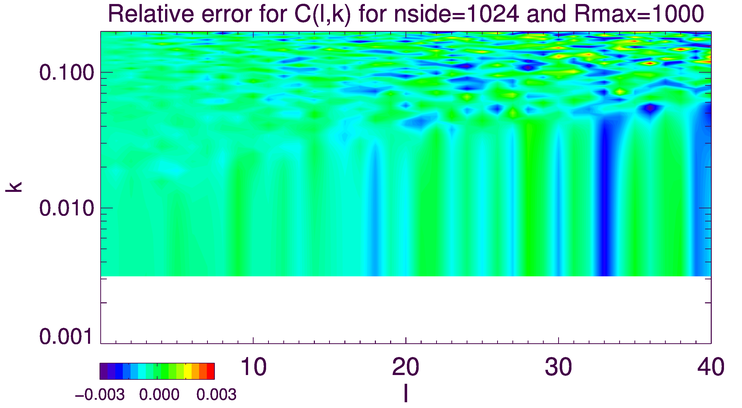} 
		\includegraphics[width=8.5cm]{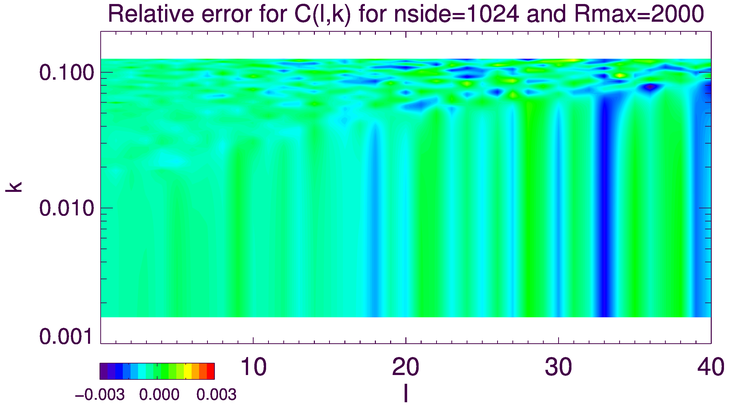}\quad}\\
	\parbox[b]{15cm}{
	 	\caption{Relative error on the power spectrum in mode-scale space $C(l,k)$ ($k$ is in ${\rm hMpc^{-1}}$). We compare the original formulation of spherical Fourier-Bessel decomposition with the reverse formulation, testing $n_{side}=512, 1024$ (rows) and $R=1000, 2000$ (columns). Only a few zones (white spots) are outside the range $[-0.3\%,+0.3\%]$.}
		\label{fig:kspace}
		}
\end{figure*}

\section{The 3DEX library}
\label{sec:3DEX}

The 3DEX library requires the HEALPix package (v2.12 or later) and the CFITSIO library. 3DEX may either by installed with an HEALPix-like procedure (\textit{configure} and \textit{make} commands) or using CMake. The Fortran modules, the 3DEX dynamic library and the related executables will be created in the relevant directories (see README file for more information).

In addition to the numerical procedures required to compute Fourier-Bessel coefficients, various other routines are provided in the library, such as those converting redshift to comoving distance,  computing spherical Bessel functions and their zeros, reading and writing 3D structures ($f_{lmn}$ and $C_{ln}$), or reconstructing radial maps from Fourier-Bessel coefficients.

Three executable programmes are generated:
\begin{itemize}
	\item \textit{survey2almn} performs the spherical Fourier-Bessel decomposition (reverse method) of a discrete survey with input parameters $l_{max}$, $n_{max}$, $r$ and $n_{side}$. Outputs are the $f_{lmn}$ coefficients and the power spectrum.
	\item \textit{survey2almn\_interactive} is very similar to the previous programme, but converts redshift values into comoving distance before performing the spherical Fourier-Bessel decomposition. In particular, the routine takes a .txt file as input, taking into account parameters on the cosmology and on the decomposition.
	\item \textit{almn2rmap} extracts the $f_{lmn}$ coefficients from a FITS file and reconstructs the field (HEALPix map) at a given radius. Inputs are the resolution, the radius and summation limits $l_{max}$ and $n_{max}$, which allows one to reconstruct several maps at different scales and resolutions.
\end{itemize}
The corresponding calls are given by the examples below. \\

{\tt $>$ survey2almn survey\_thetaphir.dat almn.fits cln.fits 20 20 256 2000.0},\\

where survey\_thetaphir.dat is a survey with columns representing $\theta, \phi, r$, and the keywords correspond to values of $\ell, n, n_{side}, R$. The output is both the coefficient values (almn.fis) and the Fourier-Bessel spectrum (cln.fits).\\

{\tt $>$ survey2almn\_interactive parameters.txt},\\

where parameters.txt is an external file containing input parameters for the survey and the cosmology (which allows for more flexible use). Finally, for the map reconstruction, we can use:\\

{\tt $>$ almn2rmap almn.fits map.fits 400.0 256 10 10 2000.0},\\

where the keywords correspond to $r_{max}$, $n_{side}$, $\ell$, $n$, $R$.

\section{Conclusion}
\label{sec:Conclusion}

High-precision cosmology from galaxy and weak lensing surveys will require the analysis of 3D data in spherical coordinates, a situation for which spherical Fourier-Bessel decomposition is most suited. Current methods will be inadequate for future planned cosmological surveys, which will provide for example galaxy surveys with billions of galaxies, compared to millions today.

We have reviewed the \emph{forward} or SHB (spherical Harmonic-Bessel) formalism of the spherical Fourier-Bessel decomposition which first calculates the tangential, then the radial decomposition.   We also introduced the \emph{reverse} or SBH (spherical Bessel Harmonic) formalism that inverses this order. Only the latter approach can take advantage of existing fast codes for the calculation of tangential modes. (To do the same, the former would require a a new voxelisation scheme.)

We presented a public code 3DEX (3D EXpansions) for the fast calculation of Fourier-Bessel coefficients and spectra, which uses the HEALPix pixelisation scheme for calculating the tangential modes. The 3DEX code is based on the \emph{reverse}/SBH formulation of the Fourier-Bessel decomposition. 

We tested the 3DEX code on linear and quasi-linear scales ($\ell<50$ and $k_{\ell n}<0.2{\rm hMpc^{-1}}$) using the Horizon halo simulation for redshifts $z<0.3$.  For $n_{side} =1024$ the 3DEX method for calculating the power spectrum $C(\ell,k)$ is accurate to 0.3\% on these scales.

For surveys with $<10$ million galaxies, computation time is reduced by a factor 4-12 depending on the desired scales and accuracy. For larger surveys the gain in time will be even greater. Finally, the use of the 3DEX code is not restricted to cosmological calculations, and can be used in any other discipline that requires a spherical Fourier-Bessel analysis of 3D data.

\begin{acknowledgements}
The authors are grateful to Ofer Lahav, Pirin Erdo\u{g}du and Fran\c cois Lanusse for useful discussions regarding the spherical Fourier-Bessel theory, as well as to Romain Teyssier and Nicolas Clerc for access and help using the Horizon simulation. The authors thank Yves Revaz for computational help at EPFL. The 3DEX library uses Healpix software \citep{healpix:2002,gorski:2004by}. This work is supported by the European Research Council grant SparseAstro (ERC-228261). This research is in part supported by the Swiss National Science Foundation (SNSF).
\end{acknowledgements}

\appendix
\label{sec:Appendices}

\section{Normalisation and discrete radial spectrum}

The basis functions $kj_l(kr)Y_{lm}(\theta,\phi)$ form a set of eigenfunctions of the Laplacian operator in spherical coordinates. In particular, these functions are orthonormalised in the continuous case, thanks to the orthogonality relation 
\begin{eqnarray}
	\nonumber \int d\Omega dr \ r^2 j_l(kr) j_{l'}(k' r) Y_{lm}(\theta,\phi) Y^{*}_{l'm'}(\theta,\phi) 
	\\ = \frac{\pi}{2kk'}\delta^D(k-k')\delta^K_{ll'}\delta^K_{mm'} \label{closure1},
\end{eqnarray}
\citep{hankel1} where $\delta^K$ is Kronecker's delta notation and $\delta^D$ Dirac's function.

{A common approach to simplify the problem is to assume some boundary conditions for the field $f$. Different conditions have been explored in the literature \citep{Fisher:1995,Heavens:1995}, including potential or gradient continuity. In this paper, we used a condition that derives from the classical formulation of the discrete spherical Bessel transform}: space is assumed to be finite and limited to a sphere of radius $R$. In this case, the spherical Bessel functions are not normalised and the boundary effect leads to a discrete spectrum $\{ k_{ln} \}$. The Fourier-Bessel coefficients become a set $f_{lmn} = f_{lm}(k_{ln})$, and the complete description of the field in the so-called Fourier-Bessel basis \citep{BQ1} is summarised in equation \ref{fbreconstr}.

As a consequence, a natural choice for the boundary condition is to impose the field to vanish at $r=R$ \citep{cmbbox}, which constrains the Bessel functions and generates the radial spectrum  $\{ k_{ln} \}$ such that, for all $l$ and $n$,
\begin{eqnarray}
	j_{l}(k_{ln}R) = 0.
\end{eqnarray}
If $q_{ln}$ denotes the $n$-th root of $j_l(z)$, the closure relation of the Bessel basis is 
\begin{eqnarray}
	\int_0^1dz \ z^2j_l(q_{ln}z)j_{l'}(q_{l'n'}z)  \nonumber \\
	= \frac{1}{2}[j_{l+1}(q_{ln})]^2 \delta_{ll'}\delta_{nn'}   ,
\end{eqnarray}
which gives  with $k_{ln} = q_{ln}/R$,
\begin{eqnarray} 
	\int_0^Rdr \ r^2 k_{ln}k_{l'n'}j_l(k_{ln}r)j_{l'}(k_{l'n'}r) \nonumber \\
	 =  \frac{k^2_{ln}[j_{l+1}(q_{ln})]^2}{2R^{-3}} \delta_{ll'}\delta_{nn'}.
\end{eqnarray}
The discrete spectrum is thus fixed by the zeros of the spherical Bessel functions.  We obtain the normalisation coefficients $\kappa_{ln}$ \citep{Fisher:1995}
\begin{eqnarray}
	\kappa_{ln}^{-1} = \frac{R^{3}}{2}[k_{ln}j_{l+1}(k_{ln}R)]^2,
\end{eqnarray}
which are used for field reconstruction (equation \ref{fbreconstr}).

Other approaches are possible to tackle boundary conditions in radial space, notably those imposing potential continuity at $r=R$ \citep{Fisher:1995}. Then, the discrete spectrum $k'_{ln}$ is such that 
\begin{eqnarray}
	j_{l-1}(k'_{ln}R) = 0,
\end{eqnarray}
and normalisation constraint becomes
\begin{eqnarray}
	{\kappa'}^{-1}_{ln} = \frac{R^{3}}{2}[k_{ln}j_{l}(k_{ln}R)]^2.
\end{eqnarray}

\section{Angular masks}

3DEX takes into account optional angular masks under the form of either an equatorial cut or an input all-sky FITS map. 

In the first case, supplying $\theta_{cut}$ defines the latitude (in degrees) of a straight symmetric cut around the equator. Pixels located within that cut ($l=cos(\theta_{cut})$) are ignored.

In the second case, the supplied mask must be an HEALPix map (ring ordering) of $N_{pix}$ pixels at resolution $n_{side}$ (which must be identical to Fourier-Bessel resolution parameter)
\begin{eqnarray}
	 \left\{w(\boldsymbol{\gamma}_q)\right\}_{q=1,\dots,N_{pix}}.
\end{eqnarray}

In the {forward method}, the first step is to apply the mask to each discrete shell before performing the spherical Harmonics Transform. Hence for the $i$th shell, field $f$ is weighted by the mask at each pixel
\begin{eqnarray}
	f'(r_i,\boldsymbol{\gamma}_q) = w(\boldsymbol{\gamma}_q) f(r_i,\boldsymbol{\gamma}_q).
\end{eqnarray}
The Fourier-Bessel coefficients are obtained after performing SH and SB transforms. 

In the reverse method, the first step is still the spherical Bessel Transform, which gives a set of $n_{max}$ HEALPix maps $f(k_{ln},\boldsymbol{\gamma}_q)$. The mask is then applied to each of these maps
\begin{eqnarray}
	f'(k_{ln},\boldsymbol{\gamma}_q) = w(\boldsymbol{\gamma}_q) f(k_{ln},\boldsymbol{\gamma}_q).
\end{eqnarray}
and the modified spherical Harmonics Transform gives the final $f_{lmn}$ coefficients.


\bibliographystyle{aa}
\bibliography{biblio}

\begin{thebibliography}{30}
\expandafter\ifx\csname natexlab\endcsname\relax\def\natexlab#1{#1}\fi

\bibitem[{{Abramo} {et~al.}(2010){Abramo}, {Reimberg}, \& {Xavier}}]{cmbbox}
{Abramo}, L.~R., {Reimberg}, P.~H., \& {Xavier}, H.~S. 2010, PRD, 82, 043510

\bibitem[{{Annis} {et~al.}(2005){Annis}, {Bridle}, {Castander}, {Evrard},
  {Fosalba}, {Frieman}, {Gaztanaga}, {Jain}, {Kravtsov}, {Lahav}, {Lin},
  {Mohr}, {Stebbins}, {Walker}, {Wechsler}, {Weinberg}, \& {Weller}}]{DES:2005}
{Annis}, J., {Bridle}, S., {Castander}, F.~J., {et~al.} 2005, astro-ph/0510195

\bibitem[{Baddour(2010)}]{hankel1}
Baddour, N. 2010, J. Opt. Soc. Am. A/Vol. 27, No. 10

\bibitem[{Binney \& Quinn(1991)}]{BQ1}
Binney, J. \& Quinn, T. 1991, -

\bibitem[{{Castro} {et~al.}(2005){Castro}, {Heavens}, \& {Kitching}}]{weak3d}
{Castro}, P.~G., {Heavens}, A.~F., \& {Kitching}, T.~D. 2005, \prd, 72, 023516

\bibitem[{{Crittenden}(2000)}]{igloo2}
{Crittenden}, R.~G. 2000, Astrophysical Letters Communications, 37, 377

\bibitem[{{Crittenden} \& {Turok}(1998)}]{igloo}
{Crittenden}, R.~G. \& {Turok}, N.~G. 1998, astro-ph/9806374

\bibitem[{Doroshkevich {et~al.}(2008)Doroshkevich, Naselsky, Verkhodanov,
  Novikov, Turchaninov, Novikov, Christensen, \& Chiang}]{glesp}
Doroshkevich, A., Naselsky, P., Verkhodanov, O., {et~al.} 2008,
  arXiv:astro-ph/0305537

\bibitem[{{Erdo{\u g}du (a)} {et~al.}(2006)}]{Erdogdu:2005wi}
{Erdo{\u g}du (a)}, P. {et~al.} 2006, \mnras, 368, 1515

\bibitem[{{Erdo{\u g}du (b)} {et~al.}(2006){Erdo{\u g}du (b)}, {Lahav},
  {Huchra}, {Colless}, {Cutri}, {Falco}, {George}, {Jarrett}, {Jones}, {Macri},
  {Mader}, {Martimbeau}, {Pahre}, {Parker}, {Rassat}, \&
  {Saunders}}]{Erdogdu:2006dv}
{Erdo{\u g}du (b)}, P., {Lahav}, O., {Huchra}, J.~P., {et~al.} 2006, \mnras,
  373, 45

\bibitem[{{Fisher} {et~al.}(1995){Fisher}, {Lahav}, {Hoffman}, {Lynden-Bell},
  \& {Zaroubi}}]{Fisher:1995}
{Fisher}, K.~B., {Lahav}, O., {Hoffman}, Y., {Lynden-Bell}, D., \& {Zaroubi},
  S. 1995, \mnras, 272, 885

\bibitem[{{G{\'o}rski} {et~al.}(2002){G{\'o}rski}, {Banday}, {Hivon}, \&
  {Wandelt}}]{healpix:2002}
{G{\'o}rski}, K.~M., {Banday}, A.~J., {Hivon}, E., \& {Wandelt}, B.~D. 2002, in
  Astronomical Society of the Pacific Conference Series, Vol. 281, Astronomical
  Data Analysis Software and Systems XI, ed. {D.~A.~Bohlender, D.~Durand, \&
  T.~H.~Handley}, 107--+

\bibitem[{{G{\'o}rski} {et~al.}(2005){G{\'o}rski}, {Hivon}, {Banday},
  {Wandelt}, {Hansen}, {Reinecke}, \& {Bartelmann}}]{gorski:2004by}
{G{\'o}rski}, K.~M., {Hivon}, E., {Banday}, A.~J., {et~al.} 2005, \apj, 622,
  759

\bibitem[{{Heavens}(2003)}]{heavens3d}
{Heavens}, A. 2003, \mnras, 343, 1327

\bibitem[{{Heavens} \& {Taylor}(1995)}]{Heavens:1995}
{Heavens}, A.~F. \& {Taylor}, A.~N. 1995, \mnras, 275, 483

\bibitem[{{Huchra} {et~al.}(2011){Huchra}, {Macri}, {Masters}, {Jarrett},
  {Berlind}, {Calkins}, {Crook}, {Cutri}, {Erdogdu}, {Falco}, {George},
  {Hutcheson}, {Lahav}, {Mader}, {Mink}, {Martimbeau}, {Schneider},
  {Skrutskie}, {Tokarz}, \& {Westover}}]{2MRS}
{Huchra}, J.~P., {Macri}, L.~M., {Masters}, K.~L., {et~al.} 2011,
  astro-ph/1108.0669

\bibitem[{{Lanusse} {et~al.}(2011){Lanusse}, {Rassat}, \&
  {Starck}}]{2011arXiv1112.0561L}
{Lanusse}, F., {Rassat}, A., \& {Starck}, J.-L. 2011, astro-ph/1112.0561

\bibitem[{{Larson} {et~al.}(2011){Larson}, {Dunkley}, {Hinshaw}, {Komatsu},
  {Nolta}, {Bennett}, {Gold}, {Halpern}, {Hill}, {Jarosik}, {Kogut}, {Limon},
  {Meyer}, {Odegard}, {Page}, {Smith}, {Spergel}, {Tucker}, {Weiland},
  {Wollack}, \& {Wright}}]{wmap7}
{Larson}, D., {Dunkley}, J., {Hinshaw}, G., {et~al.} 2011, \apjs, 192, 16

\bibitem[{{Laureijs} {et~al.}(2011){Laureijs}, {Amiaux}, {Arduini},
  {Augu{\`e}res}, {Brinchmann}, {Cole}, {Cropper}, {Dabin}, {Duvet}, {Ealet},
  \& et~al.}]{2011arXiv1110.3193L}
{Laureijs}, R., {Amiaux}, J., {Arduini}, S., {et~al.} 2011, astro-ph/1110.3193

\bibitem[{{Percival} {et~al.}(2007 b){Percival}, {Cole}, {Eisenstein},
  {Nichol}, {Peacock}, {Pope}, \& {Szalay}}]{Percival:2007}
{Percival}, W.~J., {Cole}, S., {Eisenstein}, D.~J., {et~al.} 2007 b, \mnras,
  381, 1053

\bibitem[{{Rassat} \& {Refregier}(2011)}]{2011arXiv1112.3100R}
{Rassat}, A. \& {Refregier}, A. 2011, astro-ph/1112.3100

\bibitem[{{Refregier} {et~al.}(2010){Refregier}, {Amara}, {Kitching}, {Rassat},
  {Scaramella}, {Weller}, \& {Euclid Imaging Consortium}}]{Euclidsb}
{Refregier}, A., {Amara}, A., {Kitching}, T.~D., {et~al.} 2010,
  astro-ph/0810.1285

\bibitem[{{Schlegel} {et~al.}(2007){Schlegel}, {Blanton}, {Eisenstein}, \&
  et~al.}]{Schlegel:2007}
{Schlegel}, D.~J., {Blanton}, M., {Eisenstein}, D., \& et~al. 2007, in American
  Astronomical Society Meeting Abstracts, Vol. 211, American Astronomical
  Society Meeting Abstracts, 132.29--+

\bibitem[{{Schrabback} {et~al.}(2010){Schrabback}, {Hartlap}, {Joachimi},
  {Kilbinger}, {Simon}, {Benabed}, {Brada{\v c}}, {Eifler}, {Erben},
  {Fassnacht}, {High}, {Hilbert}, {Hildebrandt}, {Hoekstra}, {Kuijken},
  {Marshall}, {Mellier}, {Morganson}, {Schneider}, {Semboloni}, {van Waerbeke},
  \& {Velander}}]{Schrabback:2010}
{Schrabback}, T., {Hartlap}, J., {Joachimi}, B., {et~al.} 2010, \aap, 516, A63+

\bibitem[{{Spergel} {et~al.}(2007){Spergel}, {Bean}, {Dor{\'e}}, {Nolta}, \&
  et~al.}]{Spergel:2006hy}
{Spergel}, D.~N., {Bean}, R., {Dor{\'e}}, O., {Nolta}, M.~R., \& et~al. 2007,
  \apjs, 170, 377

\bibitem[{{Strauss} {et~al.}(1992){Strauss}, {Huchra}, {Davis}, {Yahil},
  {Fisher}, \& {Tonry}}]{IRAS}
{Strauss}, M.~A., {Huchra}, J.~P., {Davis}, M., {et~al.} 1992, \apjs, 83, 29

\bibitem[{{Tadros} {et~al.}(1999){Tadros}, {Ballinger}, {Taylor}, {Heavens},
  {Efstathiou}, {Saunders}, {Frenk}, {Keeble}, {McMahon}, {Maddox}, {Oliver},
  {Rowan-Robinson}, {Sutherland}, \& {White}}]{1999MNRAS.305..527T}
{Tadros}, H., {Ballinger}, W.~E., {Taylor}, A.~N., {et~al.} 1999, \mnras, 305,
  527

\bibitem[{{Teyssier} {et~al.}(2009){Teyssier}, {Pires}, {Prunet}, {Aubert},
  {Pichon}, {Amara}, {Benabed}, {Colombi}, {Refregier}, \& {Starck}}]{Horizon}
{Teyssier}, R., {Pires}, S., {Prunet}, S., {et~al.} 2009, \aap, 497, 335

\bibitem[{{The Planck Collaboration}(2006)}]{Planck}
{The Planck Collaboration}. 2006, astro-ph/0604069

\bibitem[{{Tyson} \& {LSST}(2004)}]{LSST}
{Tyson}, J.~A. \& {LSST}. 2004, in Bulletin of the American Astronomical
  Society, Vol.~36, American Astronomical Society Meeting Abstracts, 108.01

\end{thebibliography}

\end{document}